\documentstyle[sprocl]{article}

\input{psfig}

\def\be{\begin{equation}}
\def\ee{\end{equation}}
\def\bea{\begin{eqnarray}}
\def\eea{\end{eqnarray}}
\begin{document}

\hspace*{7cm} JLAB-THY-99-17

\hspace*{7cm} ADP-99-24/T364	\\ \\

\title{FLAVOR ASYMMETRIES IN THE PROTON AND SEMI-INCLUSIVE PROCESSES
\footnote{Talk given at the EPIC'99 (Electron-Polarized Ion Collider)
	Workshop, IUCF, April 1999.}}

\author{W. MELNITCHOUK}

\address{Jefferson Lab, 12000 Jefferson Avenue,
	Newport News, VA 23606,		\\
	and Special Research Centre for the
	Subatomic Structure of Matter,	\\
	University of Adelaide, Adelaide 5005, Australia}


\maketitle
\abstracts{
Semi-inclusive electron scattering provides a powerful tool
with which to study the spin and flavor distributions in
the proton.
Greater kinematic coverage at the proposed Electron-Polarized
Ion Collider facility will enable the valence $d/u$ ratio to
be determined at large $x$ through $\pi^\pm$ production.
At small $x$, $\pi$ production can be used to extract the
$\bar d/\bar u$ ratio, complementing existing semi-inclusive
measurements by HERMES, and Drell-Yan data from Fermilab.
Asymmetries in heavier quark flavors can also be probed by
tagging strange and charm hadrons in the final state.}

%%%%%%%%%%%%%%%%%%%%%%%%%%%%%%%%%%%%%%%%%%%%%%%%%%%%%%%%%%%%%%%%%%%%%%%%%
\section{Introduction}

Asymmetries in the proton's spin and flavor quark distributions
provide direct information on QCD dynamics of bound systems.
Difference between quark and antiquark distributions in the sea
almost universally signal the presence of phenomena which require
understanding of strongly coupled QCD, beyond the realm of
perturbation theory.
On the other hand, at extreme kinematics at $x$ near unity
the $x$-dependence of quark distributions is determined by
perturbative QCD, and can be tested by studying the asymptotic
behavior of valence quark distributions.

Most information on quark distributions in the proton to date
has come from inclusive scattering, which sums over all
charge-squared-weighted flavors.
Semi-inclusive scattering, on the other hand, in which a hadron is
detected in the final state in coincidence with the scattered lepton,
offers considerably more freedom to explore the individual quark
flavor content of the nucleon.

In the quark-parton model the cross section for producing a
hadron $h$ at a given $x$ and $z$ and large photon virtuality
$Q^2$ can be written (to leading order) as a product of a quark
distribution function, $q(x,Q^2)$, and a fragmentation function,
$D_q^h(z,Q^2)$, giving the probability of quark $q$ fragmenting
into hadron $h$ with a fraction $z$ of the quark's center of mass
energy:
\begin{eqnarray}
\label{Ndef}
N^h &\sim& \sum_q\ e_q^2\ q(x,Q^2)\ D_q^h(z,Q^2).
\end{eqnarray}
Here one assumes factorization of the scattering and hadronization
processes, which is generally true only at high energy.
Recent data from HERMES~\cite{HERMES_DU} suggests that the
fragmentation functions are, within experimental errors,
independent of $x$, and agree with previous measurements
by the EMC~\cite{EMCFRAG} at somewhat larger energies.

%%%%%%%%%%%%%%%%%%%%%%%%%%%%%%%%%%%%%%%%%%%%%%%%%%%%%%%%%%%%%%%%%%%%%%%%%
\section{Current Fragmentation Region}

The most direct way to study fragmentation is in the current
fragmentation region, in which the observed hadrons are produced
(in the photon--hadron center of mass frame) along the direction
of the current.
Here the photon transfers (hard) momentum to a single parton in
the hadron, which is then knocked out and fragments into other
hadrons by picking up $q\bar q$ pairs from the vacuum.
Because it requires only a single $q\bar q$ pair, the leading
hadrons in this region are predominantly mesons.
Furthermore, by studying the fragmentation process at large $z$,
where the knocked out quark is most likely to be contained in
the produced meson, one can obtain direct information on the
momentum distribution of the scattered quark in the proton.
(At small $z$ this information becomes diluted by additional
$q\bar q$ pairs from the vacuum which contribute to secondary
fragmentation.)

% .......................................................................
\subsection{Valence Quark Asymmetries}

The valence $d/u$ ratio contains important information about
the spin-flavor structure of the proton~\cite{ISGUR}.
In particular, its asymptotic behavior at large $x$ reflects
the mechanism(s) responsible for the breaking
of SU(2)$_{\rm spin} \times$ SU(2)$_{\rm flavor}$ symmetry.
Furthermore, there are firm predictions for this behavior from
perturbative QCD~\cite{FJ}, so that measurement of this ratio
would be an important indicator of the appropriate kinematics
at which QCD can be treated perturbatively.

So far, a direct measurement of $d/u$ has been rather difficult,
mainly because the cross sections decrease rapidly in the extreme
kinematics near $x \sim 1$.
Previous analyses have used inclusive deep-inelastic scattering
data on proton and deuteron targets to extract the $d/u$
ratio from the neutron to proton structure function ratio.
However, the neutron data suffer from the fact that nuclear effects,
even in the deuteron, become quite significant~\cite{SMEAR,NP}
at large $x$.
In particular, whether one corrects for Fermi motion only, or in
addition for binding and nucleon off-shell effects, the extracted
neutron structure functions for $x > 0.7$ can differ
dramatically~\cite{NP}.
The question is therefore how to avoid uncertainties in the
extraction procedure introduced by nuclear effects~\cite{W,SEMID}.

One possibility is to measure the relative yields of $\pi^+$
and $\pi^-$ mesons in semi-inclusive scattering in the current
fragmentation region.
At large $z$ the $u$ quark fragments primarily into a $\pi^+$,
while a $d$ fragments into a $\pi^-$, so that at large $x$ and
$z$ one has a direct measure of $d/u$.
(Although one should not be too close to $z=1$, as the fragmentation
process there may no longer be incoherent, or factorizable into a
partonic cross section and a target-independent fragmentation
function.)

The HERMES Collaboration has previously extracted the $d/u$ ratio
from the $\pi^+$--$\pi^-$ difference using both proton and bound
neutron targets~\cite{HERMES_DU}.
The advantage of using both $p$ and $n$ is that all dependence
on fragmentation functions cancels, removing any uncertainty
that might be introduced by incomplete knowledge of the
hadronization process.
For a proton target one has (dropping the explicit $Q^2$ dependence):
\begin{eqnarray}
N_p^{\pi^+} &\sim& 4 u(x)\ D(z)\ +\ d(x)\ \bar D(z),    \\
N_p^{\pi^-} &\sim& 4 u(x)\ \bar D(z)\ +\ d(x)\ D(z),
\end{eqnarray}
where $D(z) \equiv D_u^{\pi^+} = D_d^{\pi^-}$ is the
leading fragmentation function (assuming isospin symmetry),
and $\bar D(z) \equiv D_d^{\pi^+} = D_u^{\pi^-}$ is the
non-leading fragmentation function.
With the corresponding expression for a neutron, one can
construct a ratio:
\begin{eqnarray}
\label{Rnp}
R_{np}(x,z)\ \equiv\
{ N_n^{\pi^+} - N_n^{\pi^-} \over N_p^{\pi^+} - N_p^{\pi^-} }
&=& { 4 d(x)/u(x)  - 1 \over 4 - d(x)/u(x) }\ ,
\end{eqnarray}
which is then independent of the fragmentation function,
and is a function of $x$ only.

The disadvantage of extracting $d/u$ from this ratio is that
one must still account for nuclear effects in obtaining
$N_n^{\pi^\pm}$ from the proton and deuteron data.
The HERMES Collaboration~\cite{HERMES_DU} assumed that
$N_n^{\pi^\pm} = N_d^{\pi^\pm} - N_p^{\pi^\pm}$.
Beyond $x \sim 0.7$, however, the difference between the
ratios with $N_n^{\pi^\pm}$ corrected for nuclear effects
and that which is not is quite dramatic~\cite{MST_SEMI}.
Consequently a $d/u$ ratio obtained from such a measurement
without nuclear corrections could potentially give misleading
results.

\begin{figure}[h]
\hspace*{1.5cm}
\psfig{figure=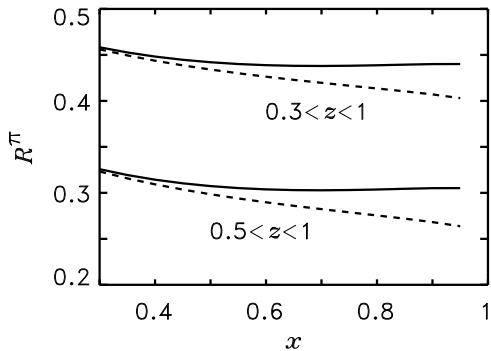,height=5.5cm}
\caption{$d/u$ ratio from semi-inclusive $\pi$ production,
	for two integration regions, $z > 0.3$ and $z > 0.5$,
	and for two asymptotic $x \rightarrow 1$ behaviors
	for $d/u$: $d/u \rightarrow 0$ (dashed) and
	$d/u \rightarrow 1/5$ (solid).}
\end{figure}

One can avoid the problem of nuclear corrections altogether
by comparing data for $\pi^+$ and $\pi^-$ production on protons
alone.
{}From the ratio of the $\pi^-$ to $\pi^+$ proton cross sections
one finds:
\begin{eqnarray}
\label{Rpi}
R^{\pi}(x,z)\ \equiv\
{ N_p^{\pi^-} \over N_p^{\pi^+} }
&=& { 4 \bar D(z)/D(z) + d(x)/u(x)
\over 4 + d(x)/u(x) \cdot \bar D(z)/D(z) }\ .
\end{eqnarray}
In the limit $z \rightarrow 1$, the leading fragmentation
function dominates, $D(z) \gg \bar D(z)$, and the ratio
$R^{\pi} \rightarrow (1/4) d/u$.
In the realistic case of smaller $z$, the $\bar D/D$ term in $R^\pi$
contaminates the yield of fast pions originating from struck primary
quarks, diluting the cross section with pions produced from secondary
fragmentation by picking up extra $q\bar q$ pairs from the vacuum.
Nevertheless, one can estimate the yields of pions using the
empirical fragmentation functions measured by HERMES~\cite{HERMES_DU}
and the EMC~\cite{EMCFRAG}.
Integrating the differential cross section over a range of $z$,
as is more practical experimentally, the resulting ratios for cuts
of $z > 0.3$ and $z > 0.5$ are shown in Fig.~1 for two different
asymptotic $x \rightarrow 1$ behaviors~\cite{ISGUR,FJ,NP}:
$d/u \rightarrow 0$ (dashed) and $d/u \rightarrow 1/5$ (solid).
Decreasing the lower limit for $z$ has the effect of raising the
cross section ratio, because of the larger integrated contribution
from non-leading fragmentation, which is more important at
smaller $z$.
Although the relative difference between the ratios for the two
asymptotic $d/u$ behaviors then becomes smaller, the absolute
difference between these remains relatively constant, and should
be measurable with the high luminosities proposed for EPIC.

% .......................................................................
\subsection{Light Sea Quark Asymmetries}

By tagging charged pions produced off protons and neutrons at
small $x$ one can also extract information on the flavor dependence
of {\em sea quarks} in the proton.
Interest in the proton's $\bar d$ and $\bar u$ distributions
has been renewed recently with the measurement by the E866
Collaboration~\cite{E866} at Fermilab of the $x$-dependence of
the $pd$ to $pp$ cross section ratio for Drell-Yan production,
which is sensitive to $\bar d/\bar u$ at small and medium $x$.
(Note that nuclear shadowing corrections, which are known to be
important in deep-inelastic scattering off the deuteron~\cite{SHAD}
at very small $x$, are relatively small in the region covered by
E866.)
Complementary information on this ratio can also be obtained in
semi-inclusive scattering by taking the ratio of the isovector
combination of cross sections for $\pi^+$ and $\pi^-$
production~\cite{LMS}:
\begin{eqnarray}
{ N_p^{\pi^+ + \pi^-} - N_n^{\pi^+ + \pi^-} \over
  N_p^{\pi^+ - \pi^-} - N_n^{\pi^+ - \pi^-} }
&=& {3 \over 5}
    \left( { u - d - \bar d + \bar u  \over  u - d  + \bar d - \bar u}
    \right)
    \left( { D + \bar D \over D - \bar D } \right) .
\end{eqnarray}
The HERMES Collaboration at DESY has in fact recently measured
this ratio~\cite{HERMES_DUBAR}, although the errors are currently
still relatively large compared with the Drell-Yan data~\cite{E866}.

As pointed out by Thomas~\cite{AWT83}, an excess of $\bar d$ quarks
in the proton arises naturally from the chiral structure of QCD,
in the form of a pion cloud.
If part of the proton's wave function can be approximated by a
$\pi^+ n$ component, a deep-inelastic probe scattering from the
virtual $\pi^+$, which contains a valence $\bar d$ quark, will
automatically lead to $\bar d > \bar u$ in the proton.
On the other hand, one can also expect the bare nucleon itself
(i.e. that which is not dressed by pions) to be asymmetric with
respect to its SU(2)$_{\rm flavor}$ sea.
As suggested long ago by Field and Feynman~\cite{FF}, the Pauli  
exclusion principle can contribute to the asymmetry on the basis
of the $u$ and $d$ valence quarks already being unequally represented
in the proton.

\begin{figure}[h]
\hspace*{1.5cm}
\psfig{figure=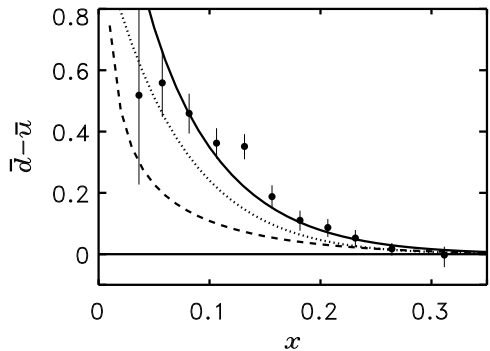,height=5.5cm}
\caption{$\bar d-\bar u$ difference arising from the pion cloud
	(dashed) and antisymmetrization (dotted) effects, and the
	combined effect (solid), compared with E866 Drell-Yan
	data~\protect\cite{E866}.}
\end{figure}

The $x$ dependence of $\bar d-\bar u$ provides much more stringent
constraints on models of the SU(2) flavor symmetry breaking in
the proton sea than earlier measurements~\cite{NMC} which only
extracted the first moment of the difference,
$\tau = \int_0^1 dx (\bar d-\bar u)$.
Analysis~\cite{DYN} of the E866 data suggests that it is rather
difficult to ascribe the whole asymmetry over the entire range
of $x$ to a single mechanism.
On the other hand, Fig.~2 illustrates that a hybrid model, where
effects of pions as well as antisymmetrization are taken into account,
gives quite a good fit to the data.

The pion cloud contributions are evaluated taking into account
both nucleon and $\Delta$ recoil states (the latter cancel
to some extent the $\bar d$ excess through coupling to the
$\Delta^{++} \pi^-$ channel).
The hadronic form factors, in the ${\cal M}$-dependent (light-cone)
dipole representation~\cite{DYN} (${\cal M}$ is the invariant mass
of the pion--baryon system), are taken to have cut-offs
$\Lambda = 1$~GeV and 1.3 GeV for the $\pi N$ and $\pi \Delta$
components, respectively.
These values, indicating a harder $\pi N \Delta$ contribution than
the $\pi N N$, are consistent with the measured values of the axial
elastic $N$ and $N \Delta$ transition form factors~\cite{AXIAL}.

The contribution from Pauli blocking, which can be parameterized as
$(\bar d~-~\bar u)^{\rm Pauli}
= \tau^{\rm Pauli} (\alpha+1) (1-x)^\alpha$,
where $\alpha$ is some large power, is based on MIT bag model
calculations~\cite{BAG} which indicate that the normalization,
$\tau^{\rm Pauli}$, can be as large as 25\%.
Phenomenologically, one finds a good fit with $\alpha \approx 14$
and a normalization $\tau^{\rm Pauli} \approx 7\%$, which is at
the lower end of the expected scale but consistent with the bag
model predictions~\cite{BAG}.
Together with the integrated asymmetry from pions,
$\tau^{\pi} \sim 0.05$, the combined value  
$\tau = \tau^{\pi} + \tau^{\rm Pauli} \approx 0.12$ is in quite
reasonable agreement with the experimental result, $0.100 \pm 0.018$
from E866.

Although the presence of an asymmetry is now clearly established,
the trend of $\bar d/\bar u$ at large $x$, where the cross sections
are smaller and errors larger, needs to be confirmed in future
experiments.
This should provide even stronger constraints on the relative
contributions of the pion cloud and antisymmetrization
contributions.

% .......................................................................
\subsection{Strange Quark Asymmetries}

Asymmetries for light sea quarks are correlated with the valence
content of the proton, so that effects of $\pi$ clouds are
difficult to distinguish from those of antisymmetrization.
Asymmetries between strange and antistrange quarks, on the other
hand, do not suffer from antisymmetrization effects, and therefore
give more direct information on the non-perturbative structure of
the nucleon sea~\cite{JI}.

Evidence for non-perturbative strangeness is currently being
sought in a number of processes, ranging from semi-inclusive
neutrino induced deep-inelastic scattering to parity violating
electron--proton scattering.
Perturbative QCD alone generates identical $s$ and $\overline s$
distributions, so that any asymmetry would have to be
non-perturbative in origin.

\begin{figure}[h]
\hspace*{1.5cm}
\psfig{figure=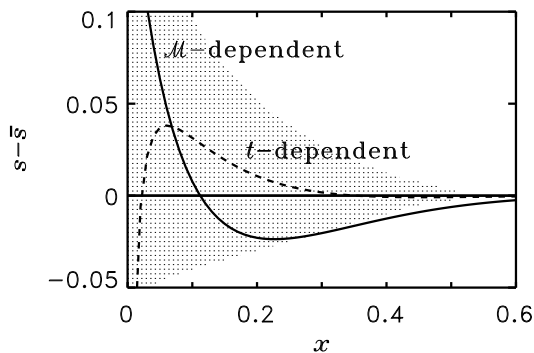,height=5.5cm}
\caption{$s-\bar s$ difference extracted from $\nu$
	deep-inelastic data~\protect\cite{CCFR} (shaded),
	with kaon cloud model predictions for ${\cal M}$-dependent
	(solid) and $t$-dependent (dashed) $KNY$ vertex functions
	with cut-off mass $\Lambda=1$ GeV.}
\end{figure}

In deep-inelastic scattering, the CCFR collaboration~\cite{CCFR}
analyzed charm production cross sections in $\nu$ and $\bar \nu$
reactions, which probe the $s$ and $\bar s$ distributions in the
nucleon, respectively.
The resulting difference $s-\bar s$, indicated in Fig.3 by the
shaded area, has been extracted from the $s/\bar s$ ratio and
absolute values of $s+\bar s$ from global data parameterizations. 
The curves correspond to chiral cloud model predictions for the
asymmetry, in which the strangeness in the nucleon is assumed
to be carried by its $KY$ ($Y=\Lambda, \Sigma$) components,
so that the $s$ and $\bar s$ quarks have quite different
origins~\cite{ST,GI}.
Because the $s$ quark originates in the $\Lambda$, its distribution
is like that of the $u$ quark in the proton, $\sim (1-x)^3$ at
large $x$.
The $\overline s$ in the kaon, on the other hand, has a much harder
shape, $\sim (1-x)$.
In the ${\cal M}$-dependent parameterization of the $KNY$ form
factor on the light-cone~\cite{MM} the $K$ distribution in the
nucleon is softer than the $\Lambda$, thereby compensating somewhat
for the harder $\bar s$ quark in the $K$.
Nevertheless, for typical form factor cut-offs of $\Lambda \sim 1$ GeV,
the softer $K$ distribution in the nucleon is not sufficient to cancel
the harder $\bar s$ quark in the $K$, so that the resulting $\bar s$
distribution in the nucleon is still bigger than $s$ at large $x$,
as in Fig.3.
As a comparison, Fig.3 also shows the asymmetry calculated for
a $t$-dependent $KNY$ form factor~\cite{MM}, for which the $K$
distribution in the nucleon is much softer than the $\Lambda$,
and does in this case overcompensate for the harder $\bar s$
distribution in $K$.
However, unlike the light-cone (${\cal M}$-dependent) forms,
the $t$-dependent form factor introduces additional problems
with gauge invariance~\cite{MM}.

\begin{figure}[h]
\hspace*{1.5cm}
\psfig{figure=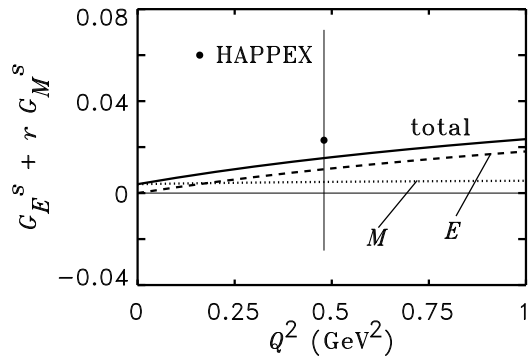,height=5.5cm}
\caption{Strange form factor combination measured by the HAPPEX
	Collaboration~\protect\cite{HAPPEX}, with kaon cloud
	predictions for the electric ($E$) and magnetic ($M$)
	contributions, and the total (solid) indicated.
	For the HAPPEX kinematics $r \approx 0.39$.}
\end{figure}

Within the same formalism as used to discuss the quark distributions,
one can also calculate the strangeness form factors of the nucleon
at low $Q^2$, which are being measured at MIT-Bates~\cite{SAMPLE}
and Jefferson Lab~\cite{HAPPEX}.
The latest data on the strange electric and magnetic form factors
is shown in Fig.4, together with the kaon cloud predictions,
using the same parameters for the $KNY$ couplings and vertex
functions as for the $s-\bar s$ difference in Fig.3.
With a soft $KNY$ form factor the contributions to both $G_E^s$
and $G_M^s$ are small and slightly positive~\cite{MM}, in good
agreement with the data.

Although the experimental results on non-perturbative strangeness
in both structure functions and form factors are still consistent
with zero, they are nevertheless compatible with a soft kaon cloud
around the nucleon.
New data on $G_{E/M}^s$ from Jefferson Lab with smaller error
bars and over a large range of $Q^2$ will hopefully provide
conclusive evidence for the presence or otherwise of a tangible
non-perturbative strange component in the nucleon.

% .......................................................................
\subsection{Intrinsic Charm}

Extending the above discussion to heavier quarks, one finds that
similar relative asymmetries (albeit smaller in magnitude) between $c$
and $\bar c$ quarks can also be generated by various non-perturbative
mechanisms.
Based on some early indications that charm production cross sections
in hadronic collisions were larger than predicted by leading order
perturbative QCD, Brodsky et al.~\cite{BROD} suggested that the
discrepancy could be resolved by introducing an intrinsic,
non-perturbative, charm component in the nucleon, associated with a
higher order, five-quark configuration in the nucleon wave function.

More recently, alternative models for non-perturbative charm have
also been proposed~\cite{NAVARRA,BM,MT_CHARM}, in which the charm
sea is assumed to arise from the quantum fluctuation of the nucleon
to a virtual $D^- \Lambda_c$ configuration, along the lines of the
$\pi$ and $K$ cloud models discussed above.
A natural prediction of this model is non-symmetric $c$ and $\bar c$
distributions, unlike the five-quark model~\cite{BROD} which assumes
$c=\bar c$.
This difference could be tested by comparing charged current
events in $e^+ p$ and $e^- p$ scattering~\cite{MT_CHARM}.

\begin{figure}[h]
\hspace*{1.5cm}
\psfig{figure=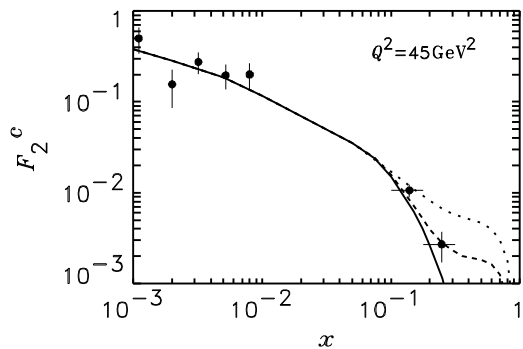,height=4.5cm}
\caption{Charm structure function calculated through the interpolating
	scheme~\protect\cite{SMT} with different amounts of intrinsic
	charm: 0\% (solid), 0.4\% (dashed) and 1\% (dotted).
	The data are from the EMC~\protect\cite{EMC}.}
\end{figure}

The most direct information on charm in the nucleon still comes
from charm production in deep-inelastic scattering, parameterized
by the charm structure function, $F_2^c$.
Since $F_2^c$ receives contributions from both perturbative and
non-perturbative charm, to unambiguously isolate the latter one
needs a reliable method of describing perturbative charm generation
over a large range of $Q^2$, including effects of heavy quark
masses and thresholds.
Such a scheme was developed recently by Steffens et al.~\cite{SMT}
to smoothly interpolate between the large-$Q^2$ limit in which
massless evolution is applicable, and the low-$Q^2$ limit where
the photon--gluon fusion process is appropriate, and at the same
time allow for the presence of non-perturbative charm.

In Fig.5 we show $F_2^c$ calculated using the interpolating
scheme~\cite{SMT} together with different amounts of intrinsic
charm from the quantum fluctuation model~\cite{MT_CHARM},
compared with data at large $x$ from the EMC~\cite{EMC}.
The EMC data appear to rule out intrinsic charm at the 1\%
level, although non-perturbative contributions of order 0.4\%
are still allowed.
More data on the semi-inclusive production of charm at large $x$
would be very valuable in settling the question of the level at
which intrinsic charm is present in the nucleon.

%%%%%%%%%%%%%%%%%%%%%%%%%%%%%%%%%%%%%%%%%%%%%%%%%%%%%%%%%%%%%%%%%%%%%%%%%
\section{Target Fragmentation Region}

While the hadrons produced in the current fragmentation region
are predominantly mesons, the baryon yield in the backward
hemisphere of the photon--target center of mass frame, or the
target fragmentation region, is observed to be higher by about
one order of magnitude.
After a collision with a high-energy photon, the remnants of the
incident hadron, when viewed in the target rest frame, usually
move with small momentum.
This makes identification of target fragments in fixed target
experiments problematic.
On the other hand, a collider set-up such as that envisioned
for EPIC, with a high luminosity beam and polarized hadrons,
would make the task of studying properties of target fragments
considerably easier.

Measurement of baryon fragments in semi-inclusive scattering from
hydrogen can provide clear indications of the mechanism underlying
the fragmentation process.
In fact, the nucleon's meson cloud gives rise to rather
characteristic fragmentation distributions in comparison
with the predictions of parton hadronization models, which
are significantly enhanced when initial and final state
polarization effects are included~\cite{TFR,ZPA}.

% .......................................................................
\subsection{Kinematics of Target Fragmentation}

Consider the semi-inclusive production of a polarized baryon $B$
(with momentum $P$) from a polarized proton (momentum $p$),
$e \vec p \rightarrow e' \vec B X$ (the electron can be unpolarized).
The four-momentum transfer squared between the initial and final
hadrons is\ 
$t = ( -p_T^2 - (1-\zeta) (M_B^2 - M^2 \zeta) )/ \zeta$,
where $\zeta = p \cdot q / P \cdot q$ is the light-cone fraction
of the proton's momentum carried by the produced baryon and $p_T$
its transverse momentum.
The requirement that the transverse momentum squared of the
baryon be positive leads to a kinematic upper limit on $t$,
namely $t_{max} = -(1-\zeta) (M_B^2 - M^2 \zeta) ) / \zeta$.

For polarized scattering, the proton polarization will be defined
to be parallel to the photon direction, and the spin of the
produced baryon quantized along its direction of motion.
We focus on production of polarized $\Delta^{++}$ baryons,
rather than for example nucleons, which will reduce backgrounds
due to $\Delta \rightarrow N$ decays.
Experimentally, the polarization of the produced $\vec\Delta^{++}$
can be reconstructed from the angular distribution of its decay
products.
For the case of $B = \Lambda$, which is relevant when studying
the distribution of strangeness in the proton, the polarization
of the $\Lambda$ hyperon will be easier to identify since the
$\Lambda$ is self-analyzing.

% .......................................................................
\subsection{Dynamics of Non-perturbative Sea Generation}

In the pion cloud model the differential cross section is determined
by the $\pi N \Delta$ vertex function, the pion structure function,
and by the amplitude, $T^{s_\Delta s_N}(t)$, for a nucleon of spin
$s_N$ to emit a pion with four-momentum squared $t$, leaving behind
a $\Delta$ with spin $s_\Delta$.
Because it is emitted collinearly (in the target rest frame) with
the pion, production of $\Delta$ baryons with helicity $\pm 3/2$
is forbidden.
The yield of spin projection $\pm 1/2$\ states, however,
is given by:
\begin{eqnarray}
{\cal T}^{ +{1 \over 2}\ \pm{1 \over 2} }(t)
&=& { 1 \over 12 M_{\Delta}^2 }
    \left[ (M - M_{\Delta})^2 - t \right]\
    \left[ (M + M_{\Delta})^2 - t \right]^2\   
    (1 \pm \cos\alpha)
\end{eqnarray}
where $\alpha$ is the angle (in the target rest frame) between
the produced baryon and the direction of the photon.
With production of $\Delta$ baryons limited to forward angles,
the factor $(1 \pm \cos\alpha)$ associated with the final state
polarization significantly suppresses the $s_\Delta = -1/2$
yield relative to that of $s_\Delta = +1/2$ final states.
The spectrum should therefore show strong correlations between
the polarizations of the incident proton ($s_N=+1/2$) and the
$\Delta^{++}$ in the pion exchange model.

A competing process to one pion exchange will be uncorrelated
spectator (target remnant, or ``diquark'' $qq$) fragmentation.
One can estimate the importance of this within the parton model
framework~\cite{TFR}, in which the differential cross section
factorizes into a product of spin-dependent quark distributions
and spin-dependent ``diquark'' fragmentation functions.

\begin{figure}[h]
\centering{
\begin{picture}(-30,150)(180,0)
\psfig{figure=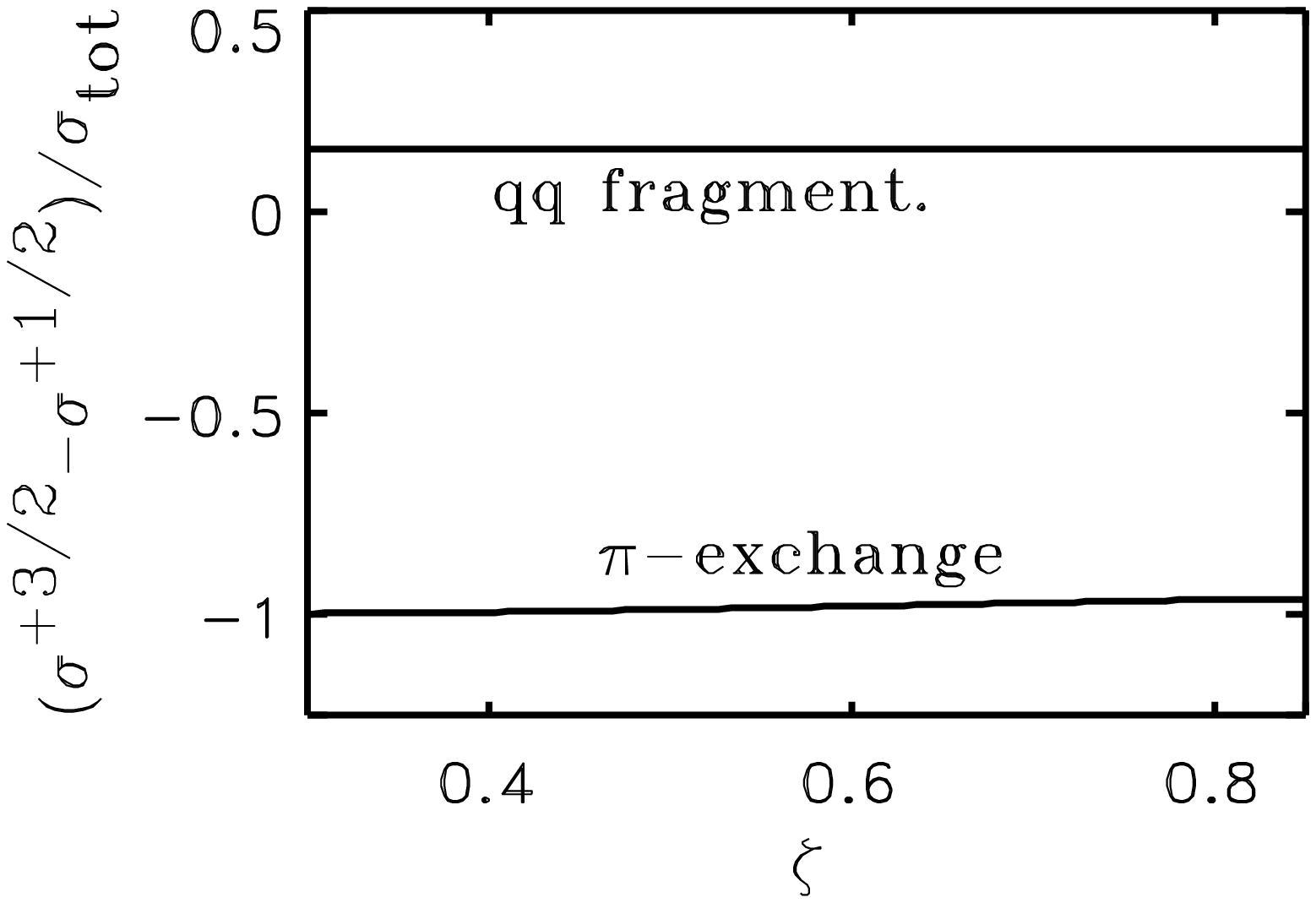,height=3.8cm}
\put(15,0){\psfig{figure=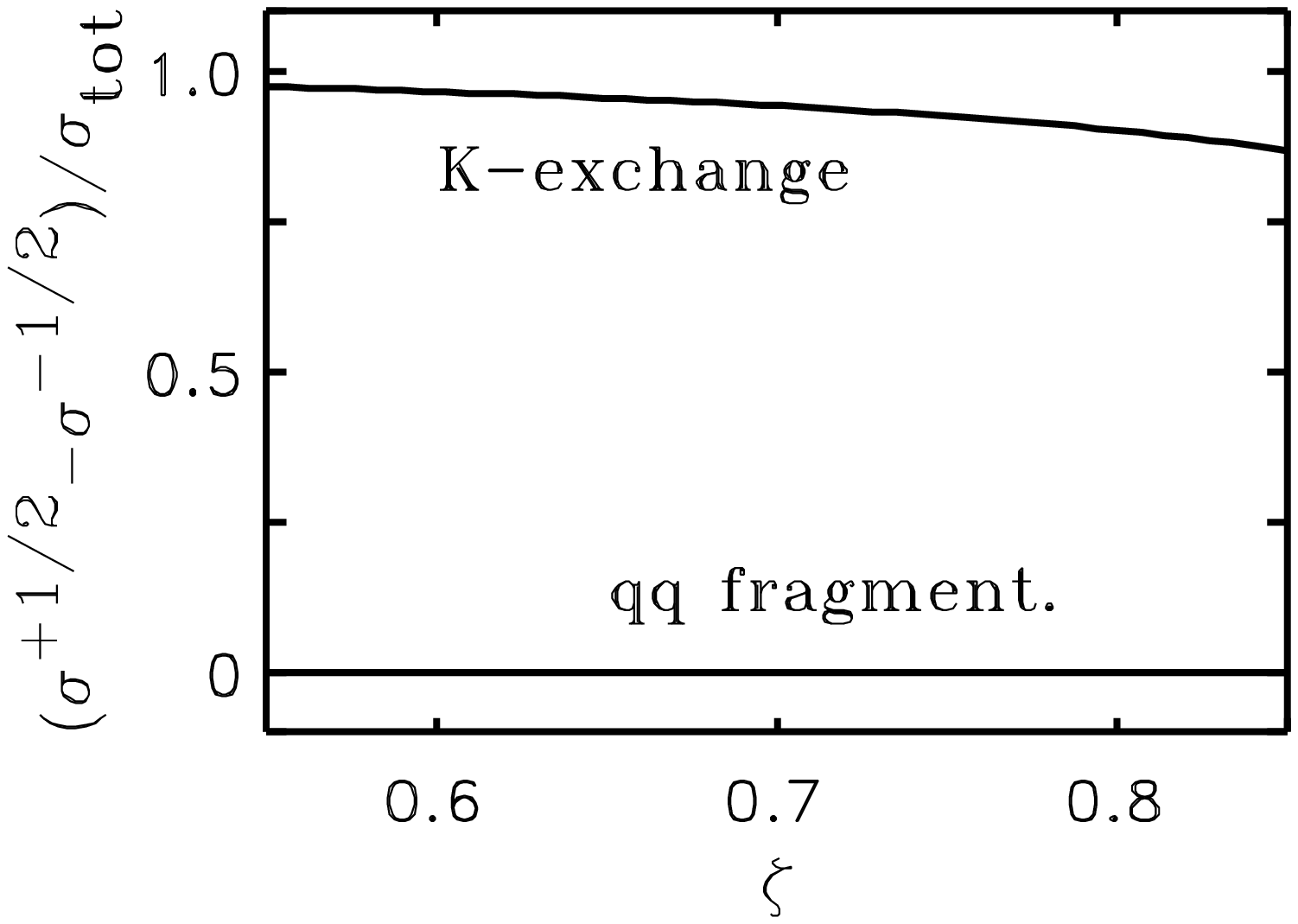,height=3.8cm}}
\end{picture}}
\caption{(a) Polarization asymmetry for $\Delta^{++}$ production in
	the $\pi$ exchange (solid) and $qq$ fragmentation (dashed)
	models.
	(b) Asymmetry for $\Lambda$ production in the $K$ exchange
	(solid) and $qq$ fragmentation (dashed) models.}
\end{figure}

In Fig.6 we show the difference $\sigma^{+3/2} - \sigma^{+1/2}$,
where $\sigma \equiv Q^2 d^3\sigma / dx dQ^2 d\zeta $,
as a fraction of the total spin-averageed cross section,
$\sigma_{\rm tot}$, for the two models at typical kinematics:
$x = 0.075$, $Q^2 = 4$ GeV$^2$ and $\sqrt s \approx 8$ GeV.
For the pion exchange model the form factor cut-off is
$\Lambda = 0.8$ GeV (giving an average number density of the
$\pi\Delta$ component\ $< n >_{\pi\Delta} \approx$ 2\%),
although the ratio is almost insensitive to the size of the
form factor cut-off.
For the $qq~\rightarrow~\Delta$ fragmentation functions we use
empirical information on unpolarized $qq$ fragmentation
together with SU(6) relations for the spin dependence~\cite{TFR}.
The resulting $\zeta$ distributions are almost flat, but
dramatically different for the two models.
In particular, the pion model predicts a large {\em negative}
asymmetry, while the asymmetry in the $qq$ fragmentation model
is small and {\em positive}.

Similarly, a kaon cloud model predicts the initial proton and
recoil $\Lambda$ polarizations to be very strongly correlated,
so that the asymmetry in Fig.6 for the cross sections is almost
unity.
The $K$ exchange ratios are very similar to the $\pi$ exchange
results, indicating the similar spin transfer dynamics inherent
to the meson cloud picture of the nucleon.
This is in strong contrast with the expectation from
$qq \rightarrow \Lambda$ diquark fragmentation, in which
the initial--final state spin correlation is much weaker.
In fact, since the probabilities to form a $\Lambda^{\uparrow}$
and $\Lambda^{\downarrow}$ from an uncorrelated $qq$ pair are
equal, in the leading fragmentation approximation the asymmetry
will be zero.
Of course, SU(6) symmetry breaking effects, as well as non-leading
fragmentation contributions, will modify this result,
as will contributions from the production and decay of
$\Sigma^{0 \uparrow\downarrow}$ hyperons.
However, the qualitative result that the asymmetry is small
should remain true.

Of course the model curves in Fig.6 represent extreme cases,
in which the $\Delta$ ($\Lambda$) baryons are produced entirely
via $\pi$ ($K$) emission or $qq$ fragmentation, with no interference
effects between them.
In reality one can expect experimental asymmetries to lie somewhere
between the two limits.
The amount of deviation from the parton model curve will indicate
the extent to which the meson-exchange process is relevant.
Unlike inclusive deep-inelastic scattering, which can only be
used to place upper bounds on the pion density~\cite{AWT83},
the semi-inclusive measurements could pin down the absolute
value of $<n>_{\pi\Delta}$ ($<n>_{K\Lambda}$).

%%%%%%%%%%%%%%%%%%%%%%%%%%%%%%%%%%%%%%%%%%%%%%%%%%%%%%%%%%%%%%%%%%%%%%%%%
\section{Conclusion}

In summary, semi-inclusive scattering offers unique opportunities
for probing the spin and flavor structure of the nucleon not
available with inclusive processes.
The polarized electron--hadron collider set-up proposed at EPIC
would allow new regions of kinematics to be accessed in which 
the dynamics of strongly interacting QCD can be probed.

We have outlined a number of specific examples where measurement
of asymmetries in the nucleon's valence and sea quark distributions
can reveal hitherto hidden details of the non-perturbative structure
of the nucleon.
The ratio of the valence $d/u$ quarks at large $x$, which reflects
the mechanism of SU(2)$_{\rm spin}\times$SU(2)$_{\rm flavor}$
symmetry breaking, can be determined by tagging pions~\cite{MST_SEMI}
in the current fragmentation region at large $z$.
Existing measurements of the neutron/proton structure function ratio,
from which $d/u$ is usually extracted, are plagued by large nuclear
binding and Fermi motion effects~\cite{NP} in the deuteron in the
region $x \approx 1$.

Pion production can also be used to pin down the ratio of the
$\bar d/\bar u$ distributions at small and medium $x$.
This can provide constraints on models which attempt to describe
the non-perturbative generation of the sea, for example in terms
of a chiral cloud of mesons around the nucleon~\cite{AWT83,DYN}.
Measurements of the light antiquark asymmetry can also be correlated
with similar asymmetries for strange and even charm quarks and
antiquarks, where effects of antisymmetrization are no longer
present.

The major advantage in studying semi-inclusive processes with
a polarized collider will be in the target fragmentation region.
As well as allowing the problem of spectator fragmentation to
be tackled seriously for the first time, careful measurements
of polarized baryon remnants from polarized protons could enable
one to unambiguously establish the magnitude of a pion and kaon
cloud of the nucleon at high energies~\cite{TFR}.
While the experiments proposed are difficult, requiring all
the intensity and duty factor one can obtain at EPIC, it does
open the way for studying the non-perturbative structure of
the nucleon in previously enexplored terrain.

\newpage
%%%%%%%%%%%%%%%%%%%%%%%%%%%%%%%%%%%%%%%%%%%%%%%%%%%%%%%%%%%%%%%%%%%%%%%%%
\section*{Acknowledgments}
I would like to thank the organizers of the EPIC'99 Workshop and the
IUCF for their hospitality and support, and travel support from the
Forschungszentrum J\"ulich.
I am indebted to M. Malheiro, J. Speth, F.M. Steffens and A.W. Thomas
for their collaboration on much of the work presented here.

%%%%%%%%%%%%%%%%%%%%%%%%%%%%%%%%%%%%%%%%%%%%%%%%%%%%%%%%%%%%%%%%%%%%%%%%%

\end{document}